\documentclass{ws-ijmpd}
\usepackage{graphicx}
\begin{document}

\markboth{Raymond Y. Chiao}{``Millikan oil drops''}

%
\catchline{}{}{}{}{} %

\title{Proposed observations of gravity waves from the early Universe 
via ``Millikan oil drops''}

\author{Raymond Y. Chiao}

\address{University of California at Merced\\P. O. Box 2030\\Merced, CA 95344,USA\\
rchiao@ucmerced.edu}

\maketitle

\begin{history}
\received{Day Month Year}
\revised{Day Month Year}
\comby{Managing Editor}
\end{history}

\begin{abstract}
Pairs of Planck-mass drops of superfluid helium coated by electrons 
(i.e., ``Millikan oil drops''), when 
levitated in a superconducting magnetic trap, can be efficient quantum 
transducers between electromagnetic (EM) 
and gravitational (GR) radiation.  This leads to the possibility of a 
Hertz-like experiment, in which EM waves are 
converted at the source into GR waves, and then back-converted at the 
receiver from GR waves back into EM 
waves. Detection of the gravity-wave analog of the cosmic microwave 
background using these drops can 
discriminate between various theories of the early Universe.
\end{abstract}


\section{Forces of gravity and electricity between two electrons}

Consider the forces exerted by an electron upon another electron at a 
distance $r$ away in the vacuum. 
Both the gravitational and the electrical force obey
long-range, inverse-square laws. Newton's law of gravitation states 
that%
\begin{equation}
\left\vert F_{G}\right\vert =\frac{Gm_{e}^{2}}{r^{2}}
\label{Newton's-inverse-square-law}
\end{equation}%
where $G$ is Newton's constant and $m_{e}$ is the mass of the electron.
Coulomb's law of electricity states that%
\begin{equation}
\left\vert F_{e}\right\vert =\frac{e^{2}}{r^{2}}\text{ }
\label{Coulomb's-law}
\end{equation}%
where $e$ is the charge of the electron. The electrical force is 
repulsive,
and the gravitational one attactive.

Taking the ratio of these two forces, one obtains the dimensionless 
constant%
\begin{equation}
\frac{\left\vert F_{G}\right\vert }{\left\vert F_{e}\right\vert 
}=\frac{%
Gm_{e}^{2}}{e^{2}}\approx 2.4\times 10^{-43}\text{ .}  \label{Gm^2/e^2}
\end{equation}%
The gravitational force is extremely small compared to the electrical 
force,
and is therefore usually omitted in all treatments of quantum physics.
Note, however, that this ratio is not strictly zero, and therefore can 
in principle be amplified.

\section{Gravitational and electromagnetic radiation powers emitted by 
two
electrons}

The above ratio of the coupling constants $Gm_{e}^{2}/e^{2}$ also is 
the
ratio of the powers of gravitational (GR) to electromagnetic (EM) 
radiation
emitted by two electrons separated by a distance $r$ in the vacuum, 
when
they undergo an acceleration $a$ relative to each other. Larmor's 
formula
for the power emitted by a single electron undergoing acceleration $a$ 
is

\begin{equation}
P_{EM}=\frac{2}{3}\frac{e^{2}}{c^{3}}a^{2}\text{ .}
\end{equation}
For the case of two electrons undergoing an acceleration $a$ relative 
to
each other, the radiation is quadrupolar in nature, and the modified 
Larmor
formula is

\begin{equation}
P_{EM}^{\prime }=\kappa \frac{2}{3}\frac{e^{2}}{c^{3}}a^{2}\text{ ,}
\label{Quad-EM-Larmor}
\end{equation}%
where the prefactor $\kappa $ accounts for the quadrupolar nature of 
the
emitted radiation.\footnote{Here $\kappa =\frac{2}{15}\frac{v^{2}}{c^{2}}$, 
where 
$v$ is their relative speed and $c$ is the speed of light, if 
$v<<c$.}
Since the electron carries mass, as well as charge, and its charge and 
mass
co-move rigidly together, two electrons undergoing an acceleration $a$
relative to each other will also emit homologous quadrupolar 
gravitational
radiation according to the formula%
\begin{equation}
P_{GR}^{\prime }=\kappa \frac{2}{3}\frac{Gm_{e}^{2}}{c^{3}}a^{2}\text{ 
.}
\label{Quad-GR-Larmor}
\end{equation}%
with the same prefactor of $\kappa $. 

The Equivalence Principle demands that the lowest order of 
gravitational
radiation be quadrupolar, and not dipolar, in nature. Hence the ratio 
of
gravitational to electromagnetic radiation powers emitted by the
two-electron system is given by the same ratio of coupling constants, 
viz.,%
\begin{equation}
\frac{P_{GR}^{\prime }}{P_{EM}^{\prime 
}}=\frac{Gm_{e}^{2}}{e^{2}}\approx
2.4\times 10^{-43}\text{ .}  \label{Ratio-of-powers}
\end{equation}%
Thus it would seem at first sight to be hopeless to try and use any
two-electron system as the means for coupling between electromagnetic 
and
gravitational radiation.

\begin{figure}[ptb]
\label{2-oil-drops-in-trap}
\par
\begin{center}
\includegraphics[width=5in]{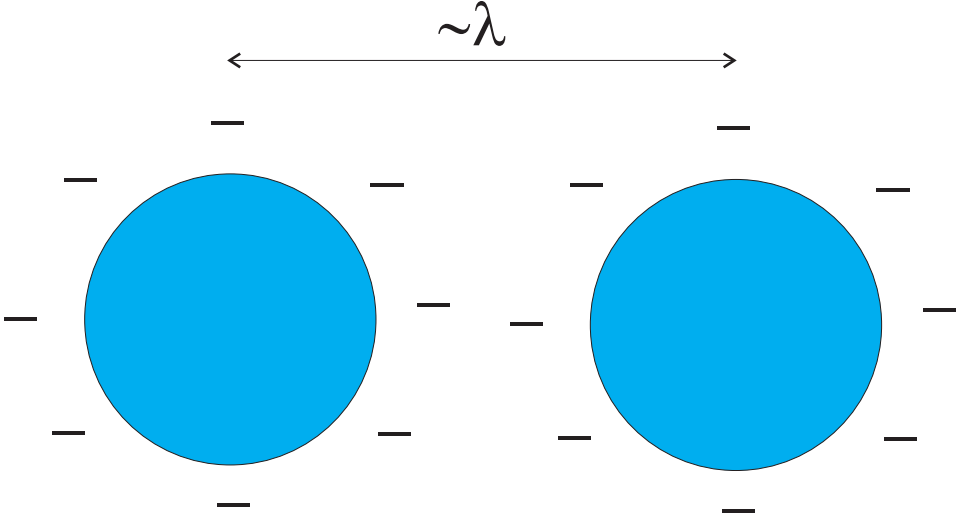}
\end{center}
\caption{A pair of levitated \textquotedblleft Millikan oil
drops\textquotedblright\ (i.e., electron-coated superfluid helium drops with
Planck-scale masses) 
in a superconducting magnetic trap, separated by around a microwave wavelength
$\lambda$.}
\end{figure}

\section{The Planck mass scale}

However, the ratio of the forces of gravity and electricity of two
\textquotedblleft Millikan oil drops\textquotedblright\ (to be 
described in more detail below; however, see Fig.~1) 
needs not be so hopelessly small.\cite{Lamb-medal}

Suppose that each \textquotedblleft Millikan oil 
drop\textquotedblright\
contains a Planck-mass amount of superfluid helium, viz.,%
\begin{equation}
m_{\text{Planck}}=\sqrt{\frac{\hbar c}{G}}\approx 22\text{ micrograms}
\label{Planck-mass}
\end{equation}%
where $\hbar $ is Planck's constant/$2\pi$, $c$ is the speed of light, and $G$ 
is
Newton's constant. Planck's mass sets the characteristic scale at which
quantum mechanics ($\hbar $) impacts relativistic gravity ($c$, $G$). 
Note
that this mass scale is mesoscopic,\cite{Kiefer-Weber} and not
astronomical, in size. This suggests that it may be possible to perform 
some
novel nonastronomical, table-top-scale experiments at the 
interface
of quantum mechanics and general relativity, which are accessible to 
the
laboratory.

The ratio of the forces of gravity and electricity between the two
\textquotedblleft Millikan oil drops\textquotedblright\ now becomes%
\begin{equation}
\frac{\left\vert F_{G}\right\vert }{\left\vert F_{e}\right\vert 
}=\frac{Gm_{%
\text{Planck}}^{2}}{e^{2}}=\frac{G\left( \hbar c/G\right) 
}{e^{2}}=\frac{%
\hbar c}{e^{2}}\approx 137\text{ .}  \label{137}
\end{equation}%
Now the force of gravity is 137 times stronger than the force of
electricity, so that instead of a mutual repulsion between these two 
charged
objects, there is now a mutual attraction between them. The sign change 
from
mutual repulsion to mutual attraction between these two 
\textquotedblleft
Millikan oil drops\textquotedblright\ occurs at a critical mass 
$m_{\text{%
crit}}$ given by%
\begin{equation}
m_{\text{crit}}=\sqrt{\frac{e^{2}}{\hbar c}}m_{\text{Planck}}\approx 
1.9%
\text{ micrograms}  \label{m_[crit]}
\end{equation}%
whereupon $\left\vert F_{G}\right\vert $ $=\left\vert F_{e}\right\vert 
$,
and the forces of gravity and electricity balance each other. This is a
strong hint that mesoscopic-scale quantum effects can lead to 
nonnegligible
couplings between gravity and electromagnetism.

The critical mass $m_{\text{crit}}$ is also the mass at which there 
occurs an
equal amount of electromagnetic and gravitational
radiation power generated upon scattering of radiation from the pair of
\textquotedblleft Millikan oil drops,\textquotedblright\ each member of 
the
pair with a mass $m_{\text{crit}}$ and with a single electron attached 
to
it. Now the ratio of quadrupolar gravitational to quadrupolar
electromagnetic radiation power is given by%
\begin{equation}
\frac{P_{GR}^{\prime }}{P_{EM}^{\prime 
}}=\frac{Gm_{\text{crit}}^{2}}{e^{2}}%
=1\text{ ,}  \label{Larmor-power-ratio}
\end{equation}%
where the prefactors of $\kappa $ in Eqs.  
(\ref{Quad-EM-Larmor}) and
(\ref{Quad-GR-Larmor}) cancel out, if the charge of the drop co-moves
rigidly together with its mass. This implies that the scattered power 
from
these two charged objects in the gravitational wave channel becomes
equal to that in the electromagnetic wave channel. However, it 
should
be emphasized that it has been assumed here that a given drop's charge 
and
mass move together as a single unit, in accordance with a 
M\"{o}%
ssbauer-like\ mode (i.e., a zero-phonon mode)\ of response to radiation
fields, which will be discussed below. This is purely quantum effect 
based
on the quantum adiabatic theorem's prediction that the system will 
remain
adiabatically, and hence rigidly, in its nondegenerate ground state 
during
perturbations arising from externally applied radiation fields.

\section{\textquotedblleft Millikan oil drops\textquotedblright\ 
described
in more detail}

Let the oil of the classic Millikan oil drops be replaced with 
superfluid
helium ($^{4}$He) with a gravitational mass of around the
Planck-mass scale, and let these drops be
levitated in a superconducting magnetic trap with Tesla-scale magnetic
fields.

The helium atom is diamagnetic, and liquid helium drops have 
successfully
been magnetically levitated in an anti-Helmholtz magnetic trapping
configuration.\cite{Weilert1996} Due to its surface tension, the 
surface of
a freely suspended, ultracold superfluid drop is atomically perfect. 
When an
electron approaches a drop, the formation of an image charge inside the
dielectric sphere of the drop causes the electron to be attracted by 
the
Coulomb force to its own image. However, the Pauli exclusion principle
prevents the electron from entering the drop. As a result, the electron 
is
bound\ to the surface of the drop in a hydrogenic ground state.
Experimentally, the binding energy of the electron to the surface of 
liquid
helium has been measured using millimeter-wave spectroscopy to be 8 
Kelvin,%
\cite{Grimes2} which is quite large compared to the milli-Kelvin 
temperature
scales for the proposed experiment. Hence the electron is tightly bound 
to
the surface of the drop.

Such a \textquotedblleft Millikan oil drop\textquotedblright\ is a
macroscopically phase-coherent quantum object. In its ground state, 
which
possesses a single, coherent quantum mechanical phase throughout the
interior of the superfluid, the drop possesses a zero circulation 
quantum
number (i.e., contains no quantum vortices), with one unit (or an 
integer
multiple) of the charge quantum number. As a result of the drop being 
at
ultra-low temperatures, all degrees of freedom other than the 
center-of-mass
degrees of freedom are frozen out, so that there results a zero-phonon 
M\"{o}%
ssbauer-like effect, in which the entire mass of the drop moves rigidly 
as a
single unit in response to radiation fields. Also, since it remains
adiabatically in the ground state during perturbations due to these 
radiation fields, the \textquotedblleft Millikan oil 
drop\textquotedblright\
possesses a quantum rigidity and a quantum dissipationlessness that are 
the
two most important quantum properties for achieving a high conversion
efficiency for gravity-wave antennas.\cite{Chiao2004}

Note that a pair of spatially separated \textquotedblleft Millikan oil
drops\textquotedblright\ have the correct quadrupolar symmetry in order 
to
couple to gravitational radiation, as well as to quadrupolar 
electromagnetic
radiation. When they are separated by a distance on the order of a
wavelength, they should become an efficient quadrupolar antenna capable 
of
generating, as well as detecting, gravitational radiation.

\section{A pair of ``Millikan oil drops'' as a transducer}

Now imagine placing a pair of levitated \textquotedblleft Millikan oil
drops\textquotedblright\ separated by approximately a microwave 
wavelength
inside a black box, which represents a quantum transducer that can 
convert
gravitational (GR) waves into electromagnetic (EM) waves. This kind of
transducer action is similar to that of the tidal force of a 
gravity
wave passing over a pair of charged, freely falling objects orbiting 
the
Earth, which can convert a GR wave into an EM\ wave. Such transducers 
are
linear, reciprocal devices.

By time-reversal symmetry, the reciprocal process, in which another 
pair of
\textquotedblleft Millikan oil drops,\textquotedblright\ converts an EM 
wave
back into a GR wave, must occur with the same efficiency as the forward
process, in which a GR wave is converted into an EM wave by a first 
pair of
\textquotedblleft Millikan oil drops.\textquotedblright\ The 
time-reversed
process is important because it allows the generation of gravitational
radiation, and can therefore become a practical source of such 
radiation.

This raises the possibility of performing a Hertz-like experiment, in 
which
the time-reversed quantum transducer process becomes the source, and 
its
reciprocal quantum transducer process becomes the receiver of GR waves 
in the far field of the source.
Room-temperature Faraday cages can prevent the transmission of EM 
waves, so
that only GR waves, which can easily pass through all classical matter 
such
as the normal (i.e., dissipative)\ metals of which standard,
room-temperature Faraday cages are composed, are transmitted between 
the two
halves of the apparatus that serve as the source and the receiver,
respectively. Such an experiment would be practical to perform using
standard microwave sources and receivers, since the scattering
cross-sections and the transducer conversion efficiencies of the two
\textquotedblleft Millikan oil drops\textquotedblright\ turns out not 
to be
too small, as will be shown below. The Hertz-like experiment would 
allow the
calibration of the \textquotedblleft 
Millikan-oil-drops\textquotedblright\
receiver for detecting the gravity-wave analog of cosmic microwave
background radiation from the extremely early Big Bang.

\section{M{\"{o}}ssbauer-like response of \textquotedblleft Millikan 
oil
drops\textquotedblright\ in a magnetic trap to radiation fields}

Let a pair of \textquotedblleft Millikan oil
drops\textquotedblright\ be levitated in a superconducting magnetic trap, 
where
the drops are separated by a distance on the order of a microwave
wavelength, which is chosen so as to satisfy the impedance-matching
condition for a good quadrupolar microwave antenna. See Fig.~1.

Now let a beam of electromagnetic waves in the Hermite-Gaussian 
TEM$_{11}$
mode,\cite{Yariv1967} which has a quadrupolar transverse field pattern 
that
has a substantial overlap with that of a gravitational plane wave, 
impinge
at a 45$^{\circ }$ angle with respect to the line joining these two 
charged
objects. As a result of being thus irradiated, the pair of 
\textquotedblleft
Millikan oil drops\textquotedblright\ will be driven into motion in an
anti-phased manner, so that the distance between them will oscillate
sinusoidally with time, according to an observer at infinity. Thus the
simple harmonic motion of the two drops relative to one another 
produces a
time-varying mass quadrupole moment at the same frequency as that of 
the
driving electromagnetic wave. This oscillatory motion will in turn 
scatter
(in a linear scattering process) the incident electromagnetic wave into
gravitational and electromagnetic scattering channels with comparable
powers, provided that the ratio of quadrupolar Larmor radiation powers 
given
by Eq.  (\ref{Larmor-power-ratio}) is of the order of unity, which will 
be
case when the mass of both drops is on the order of the critical mass 
$m_{%
\text{crit}}$ for the case of single electrons attached to each drop. 
The
reciprocal process should also have a power ratio of the order of 
unity.

The M{\"{o}}ssbauer-like response of \textquotedblleft Millikan oil
drops\textquotedblright\ will now be discussed in more detail.\ Imagine 
what
would happen if one were replace an electron in the vacuum with a 
single
electron which is firmly attached to the surface of a drop of 
superfluid
helium in the presence of a strong magnetic field and at ultralow
temperatures, so that the system of the electron and the superfluid,
considered as a single quantum entity, would form a single, macroscopic
quantum ground state. Such a quantum system can possess a sizeable
gravitational mass. For the case of many electrons attached to a 
massive
drop, where a quantum Hall fluid forms on the surface of the drop in 
the
presence of a strong magnetic field, there results a nondegenerate,
Laughlin-like ground state.

In the presence of Tesla-scale magnetic fields, an electron is 
prevented
from moving at right angles to the local magnetic field line around 
which it
is executing tight cyclotron orbits. The result is that the surface of 
the
drop, to which the electron is tightly bound, cannot undergo 
liquid-drop
deformations, such as the oscillations between the prolate and oblate
spheroidal configurations of the drop which would occur at low 
frequencies
in the absence of the magnetic field. After the drop has been placed 
into
Tesla-scale magnetic fields at milli-Kelvin operating temperatures, 
both the
single- and many-electron drop systems will be effectively frozen into 
the
ground state, since the characteristic energy scale for electron 
cyclotron
motion in Tesla-scale fields is on the order of Kelvins. Due to the 
tight
binding of the electron(s) to the surface of the drop, this would 
freeze
out all shape deformations of the superfluid drop.

Since all internal degrees of freedom of the drop, such as its 
microwave
phonon excitations, will also be frozen out at sufficiently low
temperatures, the charge and the entire mass of the \textquotedblleft
Millikan oil drop\textquotedblright\ should co-move rigidly together as 
a
single unit, in a M\"{o}ssbauer-like response to applied radiation 
fields.
This is a result of the elimination of all internal degrees of freedom 
by
the Boltzmann factor at sufficiently low temperatures, so that the 
system
stays in its ground state, and only the external degrees of freedom of 
the
drop, consisting only of its center-of-mass motions, remain.

The criterion for this M\"{o}ssbauer-like mode of 
response
of the electron-drop system is that the temperature of the system is
sufficiently low, so that the probability for the entire system to 
remain in
its nondegenerate ground state without even a single quantum of 
excitation
of any of its internal degrees of freedom being excited, is very high, 
i.e.,%
\begin{equation}
\text{Prob. of zero internal excitation}\approx 1-\exp \left( 
-\frac{E_{%
\text{gap}}}{k_{B}T}\right) \rightarrow 1\text{ as 
}\frac{k_{B}T}{E_{\text{%
gap}}}\rightarrow 0,  \label{Prob(no excitation)}
\end{equation}%
where $E_{\text{gap}}$ is the energy gap separating the nondegenerate 
ground
state from the lowest permissible excited states, $k_{B}$ is 
Boltzmann's
constant, and $T$ is the temperature of the system. Then the quantum
adiabatic theorem ensures that the system will stay adiabatically in 
the
nondegenerate ground state of this quantum many-body system during
perturbations, such as those due to weak, externally applied radiation
fields, whose frequencies are below the gap frequency 
$E_{\text{gap}}/\hbar$. By the principle of momentum conservation, since there are no
internal excitations to take up the radiative momentum, the center of 
mass
of the entire system must undergo recoil in the emission and absorption 
of
radiation. Thus the mass involved in the response to radiation fields 
is the
mass of the whole system.

For the case of a single electron (or many electrons in the case of the
quantum Hall fluid)\ in a strong magnetic field, the typical energy gap 
is
given by%
\begin{equation}
E_{\text{gap}}=\hbar\omega_{\text{cyclotron}}=\frac{\hbar 
eB}{mc}>>k_{B}T%
\text{ ,}  \label{Cyclotron-gap}
\end{equation}
an inequality which is valid for the Tesla-scale fields and 
milli-Kelvin
temperatures being considered here.

\section{Estimate of the scattering cross-section}

Let $d\sigma _{a\rightarrow \beta }$ be the differential cross-section 
for
the scattering of a mode $a$ of radiation of an incident gravitational 
wave
to a mode $\beta $ of a scattered electromagnetic wave by a pair of
\textquotedblleft Millikan oil drops\textquotedblright\ (Latin 
subscripts
denote GR waves, and Greek subscripts EM waves). Then, by time-reversal
symmetry%
\begin{equation}
d\sigma _{a\rightarrow \beta }=d\sigma _{\beta \rightarrow a}\text{ .}
\end{equation}%
Since electromagnetic and weak gravitational fields both formally obey
Maxwell's equations\cite{Wald} (apart from a difference in the signs of 
the
source density and the source current density), and since these fields 
obey
the same boundary conditions, the solutions for the modes for the two 
kinds
of scattered radiation fields must also have the same mathematical 
form. Let 
$a$ and $\alpha $ be a pair of corresponding solutions, and $b$ and 
$\beta $
be a different pair of corresponding solutions to Maxwell's equations 
for GR
and EM modes, respectively. For example, $a$ and $\alpha $ could 
represent
incoming plane waves which copropagate in the same direction, and $b$ 
and $%
\beta $ scattered, outgoing plane waves which copropagate together in a
different direction. Then for the case of a pair of critical-mass drops 
with
single-electron attachment, there is an equal conversion into the two 
types
of scattered radiation fields in accordance with Eq.  (\ref%
{Larmor-power-ratio}), and therefore%
\begin{equation}
d\sigma _{a\rightarrow b}=d\sigma _{a\rightarrow \beta }\text{ ,}
\end{equation}%
where $b$ and $\beta $ are corresponding modes of the two kinds of 
scattered
radiations.

By the same line of reasoning, for this pair of critical-mass drops%
\begin{equation}
d\sigma _{b\rightarrow a}=d\sigma _{\beta \rightarrow a}=d\sigma 
_{\beta
\rightarrow \alpha }\text{ .}
\end{equation}%
It therefore follows from the principle of reciprocity (i.e. 
time-reversal
symmetry) that%
\begin{equation}
d\sigma _{a\rightarrow b}=d\sigma _{\alpha \rightarrow \beta }.
\end{equation}

To estimate the size of the total cross-section, it is easier 
to
consider first the case of electromagnetic scattering, such as the
scattering of microwaves from two Planck-mass--scale drops, with radii $R$ 
and a
separation $r$ on the order of a microwave wavelength (but with 
$r>2R$). See Fig.~1. Let
the electrons on the \textquotedblleft Millikan oil 
drops\textquotedblright\
be in a quantum Hall plateau state, which is known to be that of a 
perfectly
dissipationless quantum fluid, like that of a superconductor. 
Furthermore,
it is known that the nondegenerate Laughlin ground state is that of a
perfectly rigid, incompressible quantum fluid.\cite{Laughlin} The two 
drops
thus behave like perfectly conducting, shiny, mirrorlike spheres, which
scatter light in a manner similar to that of perfectly elastic 
hard-sphere
scattering in idealized billiards. The total cross section for the
scattering of electromagnetic radiation from a pair of drops is 
therefore
given approximately by the geometric cross-sectional areas of two hard
spheres%
\begin{equation}
\sigma _{\alpha \rightarrow \text{all }\beta }=\int d\sigma _{\alpha
\rightarrow \beta }\simeq \text{Order of }2\pi R^{2}
\label{Geometric-X-section}
\end{equation}%
where $R$ is the hard-sphere radius of a drop.

However, if, as one might expect on the basis of classical intuitions, 
the total cross-section for the scattering of gravitational waves from 
the
two-drop system is extremely small, like that of all classical 
matter
such as the Weber bar, then by reciprocity, the total cross-section for 
the
scattering of electromagnetic waves from the two-drop system must also 
be
extremely small. In other words, if \textquotedblleft Millikan oil
drops\textquotedblright\ were to be essentially invisible to 
gravitational
radiation, then they must also be essentially invisible to electromagnetic
radiation. This would lead to a contradiction with the hard-sphere 
cross
section given by Eq.  (\ref{Geometric-X-section}), or with any other
reasonable estimate for the electromagnetic scattering cross-section of
these drops, so these classical intuitions must be incorrect.

From the reciprocity principle and from the important properties of 
quantum
rigidity and quantum dissipationlessness of these drops, one therefore
concludes that for two critical-mass \textquotedblleft Millikan oil
drops,\textquotedblright\ it must be the case that%
\begin{equation}
\sigma _{a\rightarrow \text{all }b}=\sigma _{\alpha \rightarrow 
\text{all }%
\beta }\simeq \text{Order of }2\pi R^{2}\text{ .}
\end{equation}

\begin{figure}[ptb]
\label{Veneziano}
\par
\begin{center}
\includegraphics[width=5.25in]{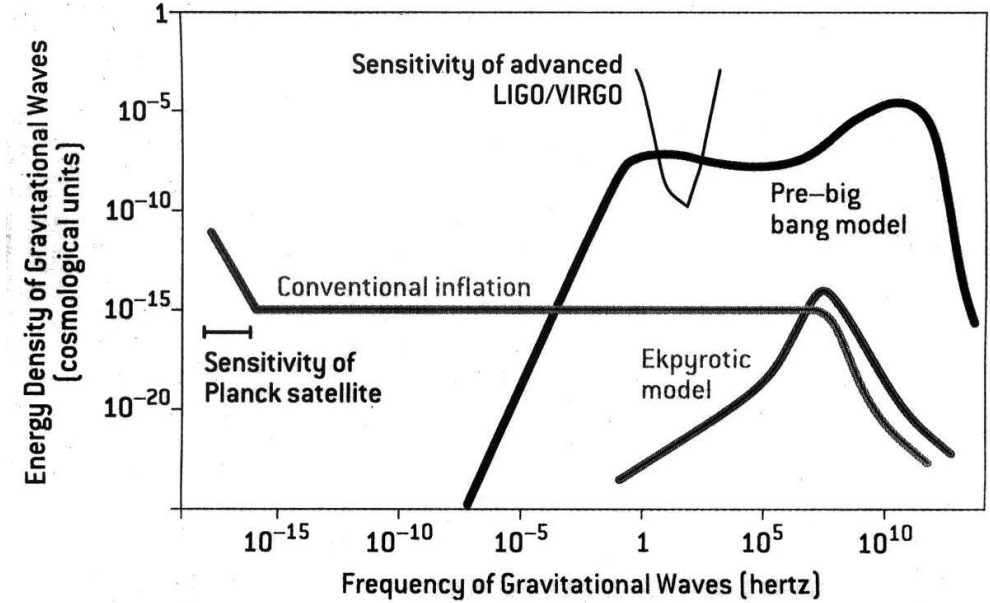}
\end{center}
\caption{Spectrum of gravity waves from the Planck-era of the Big Bang 
according to three different models. Adapted from Ref.~9.}
\end{figure}

\section{Cosmic Microwave Background in gravity waves}

An important problem in cosmology is the detection and the measurement 
of
the spectrum of gravitational radiation from the extremely early Universe,
especially around microwave frequencies. Since gravitational radiation
decouples from matter at a much earlier era of the Big Bang (i.e., the
Planck era) than electromagnetic radiation, observations of these 
primordial
gravity waves would constitute a much deeper probe of the structure of 
the
early Universe than is the case for the usual CMB.

In particular, the string-inspired pre-Big-Bang model, the ekpyrotic 
model
based on brane theory, and the conventional inflation model, give 
totally
different predictions as to the gravity-wave spectrum.\cite{Veneziano}
See Fig.~2. Observations in the radio- and microwave-frequency parts of the 
spectrum
would be decisive in determining which model (if any) is the correct 
one,
since the position of the maxima in the spectra predicted by the
pre-Big-Bang and ekpyrotic models and their strengths are strikingly
different from each other. Both models in turn yield spectra which 
differ
greatly from the spectrum predicted by the conventional inflation 
model,
which is extremely flat up to the microwave frequency range, where 
there is
a cutoff, but where there are no maxima at all.

\section*{Acknowledgments}

I would like to thank Slava Turyshev for inviting me to participate in
NASA's recent \textquotedblleft Quantum to Cosmos\textquotedblright\
conference.

\end{document}